\documentclass{aa}
\usepackage{amsmath,amssymb}
\usepackage{graphicx}
\usepackage{hyperref}

\newcommand{\fig}[1]{Fig. \ref{#1}}
\newcommand{\refsection}[1]{Section \ref{#1}}
\newcommand{\reftable}[1]{Table \ref{#1}}
\newcommand{\refeq}[1]{Eq. (\ref{#1})}

\newcommand{\BothTen}{10G10}
\newcommand{\BothPointOne}{DM0.1G0.1}
\newcommand{\BothOne}{DM1G1}
\newcommand{\DMTen}{DM10G1}
\newcommand{\DMPointOne}{DM0.1G1}
\newcommand{\GasTen}{DM1G10}
\newcommand{\GasPointOne}{DM1G0.1}

\begin{document}
\title{Probing modified gravity via the mass-temperature relation of galaxy clusters.}

\author{A. Hammami
	\inst{1}
        \and
	D. F.  Mota
	\inst{1}}

\institute{Institute of Theoretical Astrophysics, University of Oslo, P.O. Box 1029 Blindern, N-0315 Oslo, Norway\\
	    \email{amirham@astro.uio.no}\\ 
	    \email{d.f.mota@astro.uio.no}}

\abstract
{We propose that the mass-temperature relation of galaxy clusters is a prime candidate for testing gravity theories beyond Einstein's general relativity, for modified gravity models with universal coupling between matter and the scalar field. For non-universally coupled models we discover that the impact of modified gravity can remain hidden from the mass-temperature relation.
Using cosmological simulations, we find that in modified gravity the mass-temperature relation varies significantly from the standard gravity prediction of $M \propto T^{1.73}$. 
To be specific, for symmetron models with a coupling factor of $\beta=1$ we find a lower limit to the power law as $M\propto T^{1.6}$; and for f(R) gravity we compute predictions based on the model parameters. 
We show that the mass-temperature relation, for screened modified gravities, is significantly different from that of standard gravity for the less massive and colder galaxy clusters, while being indistinguishable from Einstein's gravity at massive, hot galaxy clusters.
We further investigate the mass-temperature relation for other mass estimates than the thermal mass estimate, and discover that the gas mass-temperature results show an even more significant deviations from Einstein's gravity than the thermal mass-temperature.
}

\maketitle

\section{Introduction}
Ever since the discovery that the universe is undergoing a late-time accelerated expansion \citep{acc_universe} the biggest challenge within the field has been to argue why this is happening. The two main hypothesis are that the expansion is driven by some unknown exotic component known as dark energy \citep{dark_energy}, or that the accelerated expansion is the sign that general relativity need to be modified at large scales \citep{mod_grav_review,BransDick}.

The biggest challenge of modified gravity theories is to alter the behaviour of gravity on large scales, where the accelerated expansion is observed, while leaving the gravity interaction on smaller scales unchanged, where general relativity has been tested with high precision \citep{abexp, labexp2, solarsyst,gravityprobe}. To accomplish this,  several screening mechanisms have been proposed \citep{screen1, screen2}. In this paper, we study two specific cases, the chameleon f(R)-gravity \citep{husawicki} and the symmetron scalar tensor theory \citep{SymmetronPaper}. Notice however that our results are valid for a general class of theories of modified gravity.

Modifying general relativity at cosmological scales affects structure formation. In the case of theories with a screening mechanism, the main signatures occur in the nonlinear regime and at galaxy cluster scales.  Performing dark-matter only N-Body simulations within these modified gravity theories is therefore a promising way of probing their effects \citep{fofrnbodychicago, LlinaresKnebe, Li1, Li2, Li3, ISIS, nbodycham, nbodyvainst, nbodysymmnqs, symmfofrredshift, Winther}. However, there is a major pitfall in such approach: in order to test these theories against observations one needs to compute real observables. The latter cannot be directly calculated from dark matter only simulations, since experiments only measure photons, which are in fact emitted from the baryonic matter.

This raises a major question: what observables from the simulations would be best suited for comparing to observations, in order to put stronger constrains on modified gravity theories and test Einstein's general relativity? In order to tackle such crucial problem  N-Body simulations for modified gravity theories have started to include hydrodynamics to simulate the behaviour and observables associated with baryons \citep{hammami1, hammami2, MG-Gadget,nrtwo}.  Cluster properties such as halo profiles and probability distribution functions have been computed, and lately, the gas-fraction of the galaxy clusters and power-spectra have been suggested as viable candidates \citep{hammami2, copycatLi}.

In this paper, we propose to use the mass-temperature relation of a galaxy cluster as a new and quite unique observable for testing gravity theories. We show that it can be used to set strong constraints on modified gravity theories and to test general relativity in a new region of the parameter space.

We also show that the mass-temperature relation is a very promising probe in part due to the vast amount of new, high resolution X-ray data from XMM Newton and Chandra, and also due to the quite specific signatures that different models predict. Therefore, allowing us to probe the nature of gravity at cluster scales. 

In \refsection{theory_sec} we briefly introduce the theoretical framework for our chosen modified gravity theories and the mass-temperature relation. In \refsection{observations_sec} we discuss which observations to use and the various assumptions used in the literature. In \refsection{simulation_sec} we describe our simulations and how we calculate the mass-temperature relation. In \refsection{results_sec} we show the results from our simulations and compare them to the observations from \refsection{observations_sec}. We summarize and give our final thoughts in \refsection{conclusions_sec}.

\section{Theory}
\label{theory_sec}
\subsection{Modified gravity}

The symmetron model and f(R)-gravity are both scalar-tensor theories of gravity that can be defined by the same general action
\begin{align}
\label{action}
S = \int d^4x\sqrt{-g}\left[\frac{R}{2}M_{pl} - \frac{1}{2}\partial^i\varphi\partial_i\varphi  - V(\varphi)\right] + S_m(\tilde{g}_{\mu\nu}, \varphi_i),
\end{align}
where $R$ is the Ricci scalar, $M_{pl}$ is the Planck mass, $\varphi$ is the scalar field, $V(\varphi)$ is the potential, $\psi$ are the matter fields, $g$ is the determinant of the metric tensor $g_{\mu\nu}$. The scalar field is conformally coupled to matter by the conformal factor $\tilde{g}_{\mu\nu} ) A(\varphi)^2g_{\mu\nu}$, which will result in an extra, fifth, force of the form
\begin{align}
F_{\varphi} = -\frac{A'({\varphi})}{A(\varphi)}\nabla\varphi.
\end{align}


\subsubsection{Symmetron}
The symmetron model \citep{SymmetronPaper} possess a screening mechanism that is sensitive to the local density. If the density is high, the scalar degree of freedom decouples from matter, and the fifth force becomes negligible. In regions of low density, the coupling between matter and the extra field is strong, and the fifth force reaches its maximum value. This mechanism is ensured by having a symmetric coupling function and potential, around the value $\varphi = 0$,

\begin{align}
A(\varphi) = 1 + \frac{1}{2}\left(\frac{\varphi}{M}\right)^2
\end{align}
and
\begin{align}
V(\varphi) = V_0 - \frac{1}{2}\mu^2\varphi^2 + \frac{1}{4}\lambda\varphi^4, 
\end{align}
where $M$ and $\mu$ are mass scales and $\lambda$ is a dimensionless parameter. These free parameters can be recast to parameters with a more intuitive physical interpretation
\begin{align}
\beta = \frac{M_{pl}\varphi_0}{M^2}, \\
a_{SSB}^3 = \frac{3H_0^2\Omega_m M_{pl}}{M^2\mu^2}, \\
\lambda_0^2 = \frac{2\mu^2},
\end{align}
 where $\varphi_0$ is the scalar field minimum, which vanishes in regions of high density, $H_0$ is the Hubble constant and $\Omega_M$ is the matter density parameter of the Universe. These parameters now represent
\begin{itemize}
\item $\beta$ - The strength of the scalar field, and therefore the amplitude of the fifth force.
\item $a_{SSB}$ - The expansion factor of the Universe at the time of symmetry breaking. Prior to this the density of the Universe had the fifth force permanently screened.
\item $\lambda_0$ - The range of the fifth force, in units of $Mpch^{-1}$.
\end{itemize}

With the symmetron coupling function, the fifth force becomes
\begin{align}
F_{\varphi} = -\frac{\varphi}{M^2}\nabla \varphi = 6\Omega_m H_0^2 \frac{\beta^2\lambda_0^2}{a_{SSB}^3}\tilde{\varphi}\nabla\tilde{\varphi},
\end{align}
in the last step a switch to super-comoving coordinates has been made, as detailed in \citet{hammami1, hammami2}

\subsubsection{$f(R)$-gravity}
The $f(R)$-gravity models are a set of extended gravity theories where the Einstein-Hilbert Lagrangian density $\mathcal{L}_{\rm EH} = R$ is replaced by a more general function of the Ricci scalar $f(R)$. 

The action describing the $f(R)$-gravity theories,
\begin{align}
 S = \int\sqrt{-g}\left[\frac{R+f(R)}{16\pi G} + \mathcal{L}_m\right]d^4x,
\label{husawick}
\end{align}
can be transformed to the form of the general action for scalar-tensor theories \refeq{action} using the conformal transformation
\begin{align}
A(\varphi) = \exp\left(-\frac{\beta\varphi}{M_{pl}}\right),
\end{align}
where the coupling factor is constant $\beta=\sqrt{1/6}$. 

These theories possess a so-called Chameleon screening mechanism, where the mass of the scalar field is dependent on the local density, which in turn decide the interaction range of the scalar field \citep{Chameleons}. If the density is high, the scalar degree of freedom becomes very short ranged,  while in low dense the range is large and deviations from general relativity reach its maximum value.

For this paper we will be working with the Hu-Sawicki $f(R)$ model \citep{husawicki}
\begin{align}
f(R) = -m^{2(1-n)}\frac{c_1R^n}{1 + c_2(R/m^2)^n},
\end{align}
where $m^2 = H_0^2\Omega_{m0}$ and $n$, $c_1$ and $c_2$ are free parameters. We can reduce the number of free parameters to be $n$ and $f_{R0}$ by the relations
\begin{align}
c_1 &= 6c_2\frac{\Omega_{\Lambda}}{\Omega_m},
\end{align}
and
\begin{align}
f_{R0} &= -n\frac{c_1}{c_2^2}\left(\frac{\Omega_{\Lambda}}{3(\Omega_m + 4\Omega_{\Lambda})}\right)^{n+1}.
\end{align}
The range of the scalar degree of freedom is dependent on these parameters as $\lambda_0 \propto \sqrt{1/f_{R0}}$.  

With the Hu-Sawicki $f(R)$-gravity formalism the fifth force becomes
\begin{align}
F_{\varphi} = -\frac{a^2\beta}{M_pl}\nabla\varphi.
\end{align}
Further details can be found in our previous work \citep{hammami1} or in the review by \citet{f(R)Theories}.

\subsection{Mass-temperature relation}
The virial theorem that relates the kinetic energy $T$ to the gravitational potential $U$, 
\begin{align}
2T + U = 0, \label{virial}
\end{align}
can be used to find a simple theoretical mass-temperature relation
\begin{align}
\label{theoretical_mt}
M \propto T^{3/2},
\end{align}
valid at the virial radius $r_{\rm vir}\approx r_{\rm 200c}$\footnote{$r_{\rm 200c}$ is defined as the radius where the density of a galaxy cluster is 200 times the critical density of the Universe and is generally thought to be the point where the halo is fully virialized and at hydrostatic equilibrium.}. The full expression is very complicated \citep{Lilje}, consisting of cosmological parameters and the density profile. Furthermore, studies have shown \citep{popolo} that the relation grow steeper for the low mass-temperature range, and that two separate power-laws can describe the low mass-temperature and high mass-temperature range respectively.

Observational investigations of the mass-temperature relation have been performed to test this theoretical relation. However no consensus has been reached with results ranging from substantially lower and higher than $3/2$. In this paper we aim to investigate how this relation changes within modified gravity. 

In order to test this relation, the thermal mass must be constructed using our simulation gas output, which currently consist of pressure $p$, density $\rho$  and velocity $v$.

In order to compute the thermal mass one assumes that a galaxy cluster has reached hydrostatic equilibrium at the present epoch, $z=0$, and at a radius of r$_{\rm 200 c}$ \footnote{It has been shown \citep{XMM} that given all other uncertainties involved this is a good assumption.},  we express this as
\begin{align}
\frac{dP}{dr} = -\frac{GM(r)\rho(r)}{r^2},
\end{align}
where $G$ is Newton's gravitational constant and $M(r)$ is the mass within radius $r$. Using the ideal gas relation between pressure and temperature $P_{\rm thermal} = k_Bn_{\rm gas} T_{\rm gas}$ with $\rho_{\rm gas} = \mu m_p n_{\rm gas}$ we get the mass within radius $r$ to be
\begin{align}
\label{thermalmass}
M(r) = -\frac{k_Br^2T_{\rm gas}(r)}{\mu m_pG}\left(\frac{d\ln\rho_{\rm gas}}{d\ln r} +  \frac{d\ln T_{\rm gas}}{d\ln r} \right),
\end{align}
where $\mu=0.59$ is the mean molecular weight of the gas and $k_B$ is the Boltzmann constant.

Experimentally, one calculates the thermal mass by measuring the temperature and density profiles via X-ray temperature, Sunyaev--Zel'dovich effect and surface-brightness observations \citep{TestCham, XMM}. 
Theoretically, one computes the thermal mass in the same way: temperature and density profiles can be directly obtained from our hydrodynamic and n-body simulations for the different modified gravity models.

\subsection{The $Y_X$ mass indicator proxy}
An alternative to studying the mass-temperature relation exist in the form of the $Y_X$ proxy introduced by \citet{Kravtsov06}. The proxy is defined as the product of the spectral temperature and the mass of the gas in a galaxy cluster
\begin{align}
Y_X = T_{\rm spec} M_{\rm gas}.
\end{align}
Studies show that the $Y_X$ proxy has a low scatter at high and low redshifts independent of whether the cluster is relaxed or not. In short the $Y_X$ might prove a better probe for modified gravity theories, particularly since it is not as sensitive to astrophysical uncertainties as the mass temperature relation. 

$Y_X$ is a function of the spectral temperature, while the simulations contain a gas mass weighted temperature. It is found in \citet{0507092} that it is possible to relate these temperatures to one another by a simple factor of
\begin{align}
T_{\rm spec} = 0.9 T_{\rm gas}.
\end{align} 
This result is based on observations of 12 galaxy clusters and is completely empirical, with no assumptions of theories of gravity.
 
\section{Observations}
\label{observations_sec}
The observations found in the literature can be categorized as either having well defined spatial- or spectral resolution. Historically there have been very few observations with high spatial resolution \citep{9902151} where the profiles can be directly observed. The majority of the observations need to construct the profiles using analytical and numerical models, with the most common being the isothermal $\beta$-model. We are hopeful that the increased resolution of XMM Newton and Chandra can alleviate this problem in the future.

There exist a large range of studies of the mass-temperature relation \citep{NeumannArnaud, EttoriFabian, PhD-thesis,  Nevalainen, Finoguenov, Xu2001,0212284, 0507092, X-ray_prop, XXL} in the literature, however, not all of these are readably usable for our purpose. Our simulations have derived the quantities at $r_{\rm 200c}$, we therefor require that the observations also have their quantities measured at the same radius. There exist methods for converting one set of $r_{\rm \Delta c}$ to any other value \citep{0212284}, however it require us to have a model for the density profile, a restriction we’d not like to impose on our comparisons.

In the end we chose to work with the data from \citet{X-Ray_prop} and \citet{PhD-thesis}. \citet{X-ray_prop} is presenting stacked galaxy cluster masses and temperatures from the ROSAT All-Sky Survey counting over 4000 clusters, while \citet{PhD-thesis} presents data from the ASCA cluster catalogue counting 273 clusters. 

For the $Y_X$ mass proxy we obtain data from \citet{Eckmiller11, Lovisari15}, using 26 clusters from Chandra and 82 clusters from XMM-Newton respectively. The data for the $Y_X$ mass proxy is unfortunately not available at  $r_{\rm 200c}$, but only at $r_{\rm 500c}$. To work around this we calculate the $Y_X-T$ relation at $r = 0.63 r_{\rm 200c}$, which is a good approximation of $r_{\rm 500c}$.

\section{Simulations}
\label{simulation_sec}
Our code is a modification of the ISIS code \citep{ISIS}, which in turn is a modification of the cosmological, hydrodynamic N-body code RAMSES \citep{Ramses}. ISIS implemented the $f(R)$-gravity and symmetron models to the dark matter component of RAMSES, while the current code extended the modified gravity to also work on the hydrodynamic part of RAMSES.

We run two sets of simulations; one for the symmetron models and one for the $f(R)$-gravity models. Due to consistency with previous work \citep{hammami1, hammami2} the background cosmology and box size differ in these two sets. Both sets contain $256^3$ dark matter particles. 

For the $f(R)$-gravity set we have a box size of 200 $Mpc/h_0$, with $h_0 = 0.7$,  $\Omega_{\Lambda} = 0.727$, $\Omega_{\rm CDM} = 0.227$ and $\Omega_{b} = 0.045$. The resulting dark matter particle mass is $3\times10^{10} M_{\odot}/h$.

For the symmetron set we have a box size of 256 $Mpc/h_0$, with $h_0 = 0.65$,  $\Omega_{\Lambda} = 0.65$, $\Omega_{\rm CDM} = 0.3$ and $\Omega_{b} = 0.05$. The resulting dark matter particle mass is $8.32\times10^{10} M_{\odot}/h$.

The two different cases of the $\Lambda$CDM model need to be distinguishable in the text. We denote the background $\Lambda$CDM model using the symmetron box size and background as $\Lambda$CDM$_{\rm S}$ and the one using the $f(R)$-gravity box size and background as $\Lambda$CDM$_{\rm f(R)}$.

An overview of the model parameters employed is found in \reftable{params_tab}.

\begin{table}
 \begin{center}
\caption{Overview of the model parameters for the symmetron and $f(R)$ models.}
\label{params_tab}
  \begin{tabular}{lrrr}\hline
  Symmetron models  & $\beta$ & $ a_{\rm SSB}$ & $\lambda_{\varphi}$ \\ \hline
  Sym A & 1.0 & 0.5 & 1.0 \\ 
  Sym B & 1.0 & 0.33 & 1.0 \\ 
  Sym C & 2.0 & 0.5 & 1.0 \\ 
  Sym D & 1.0 & 0.25 & 1.0 \\ \hline\hline\\\hline
  $f(R)$ models   & $ f_{R0}$&$ n$& \\ \hline
  FofR04 & $10^{-4}$&1& \\ 
  FofR05 & $10^{-5}$&1& \\ 
  FofR06& $10^{-6}$&1& \\ \hline\hline
  \end{tabular}
 \end{center}
\end{table}

\section{Results}
\label{results_sec}
Using the Amiga Halo Finder \citep{AHF} we obtain the location of all the galaxy clusters and their respective $r_{\rm 200c}$. We keep all galaxy clusters that contain at least 100 dark matter particles to ensure that all clusters we study are well above the resolution limit of our simulations. In general this leaves somewhere between 9000 to 10000 clusters per model.

All mass quantities are scaled by the dimensionless Hubble parameter and presented in units of $M_{\odot}h_0^{-1}$ and the temperature is scaled by the Boltzmann constant $k_B$ and presented in units of $keV$.

\begin{figure}
        \centering
        \includegraphics[width=0.45\textwidth]{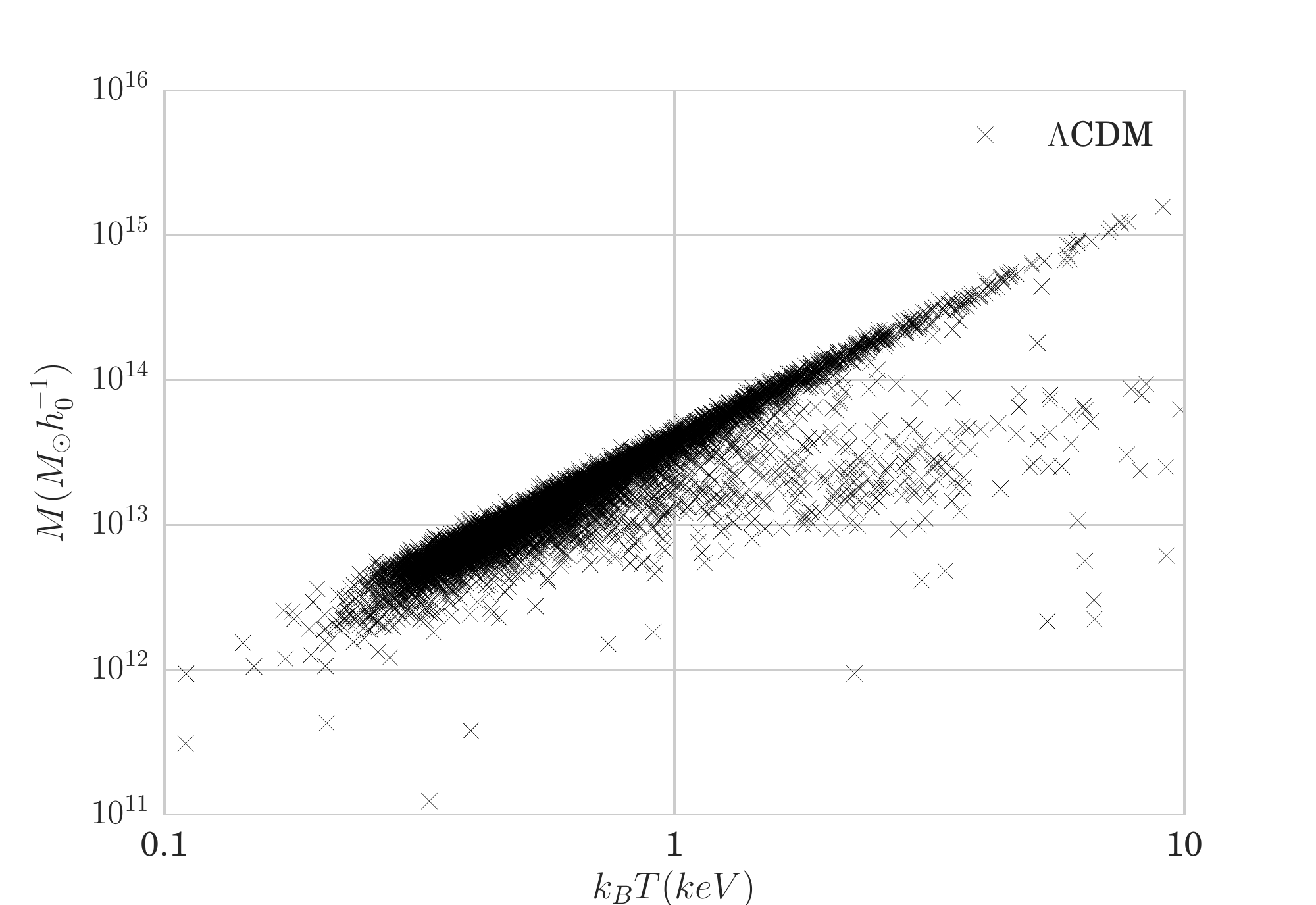}
	\vspace{-1 mm}
\caption{The raw mass-temperature data for $\Lambda$CDM$_{\rm S}$, containing all 9307 identified galaxy clusters.}
 \label{m-t-ind_lcdm}
 \end{figure}

\fig{m-t-ind_lcdm} show the raw mass-temperature plot for all the identified galaxy clusters in the $\Lambda$CDM$_{\rm S}$ model, the same plot for $\Lambda$CDM$_{\rm f(R)}$ is very similar and will not be shown.

The majority of the mass and temperature is, in general, following a near-linear relation, however with a notable amount of outliers. The outliers consist of low-mass halos with very large temperatures. Attempting to find an analytical fit to this result would be highly skewed due to the outliers. The outliers appear in all our models, both standard gravity as well as modified gravity. We assume that these are clusters that have not yet reached hydrostatic equilibrium, and that \refeq{thermalmass} is not valid for these clusters.

However, the mass-temperature relation using stacked\footnote{We stack the quantities in mass bins and then calculate the average. The bins are chosen so that the largest bin have a minimum of 15 clusters, the remaining bins contain hundreds of clusters} masses and temperatures show no signs of any outliers, which have been suppressed by the stacking process. For the remainder of the paper we will be discussing the mass-temperature relation constructed from stacked quantities.

\begin{figure}
        \centering
        \includegraphics[width=0.45\textwidth]{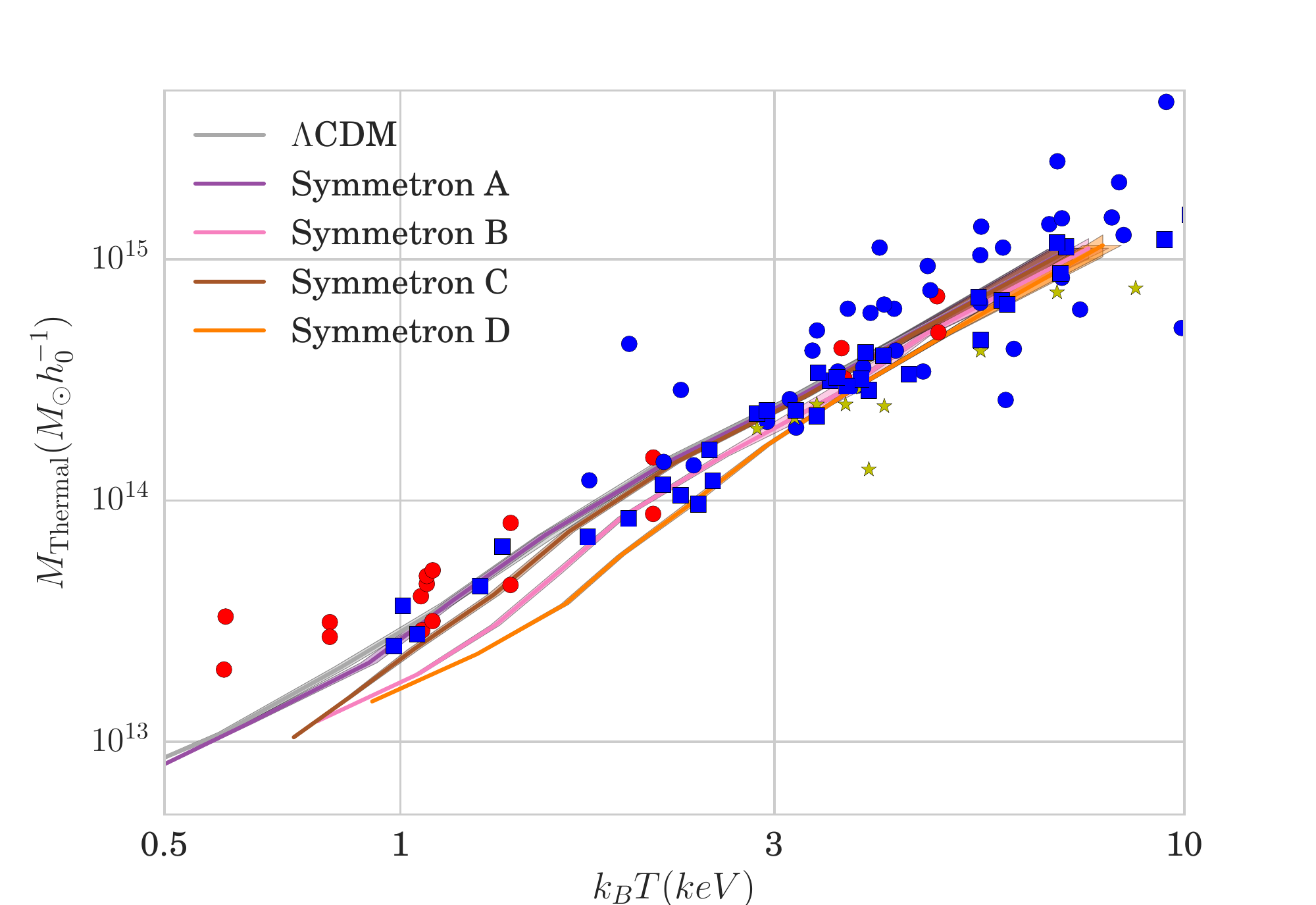}
	\vspace{-1 mm}
\caption{The mass-temperature relation for $\Lambda$CDM$_{\rm S}$ and the symmetron models, for the stacked galaxy clusters.}
 \label{m-t-symmetron}
 \end{figure}

\begin{figure}
        \centering
        \includegraphics[width=0.45\textwidth]{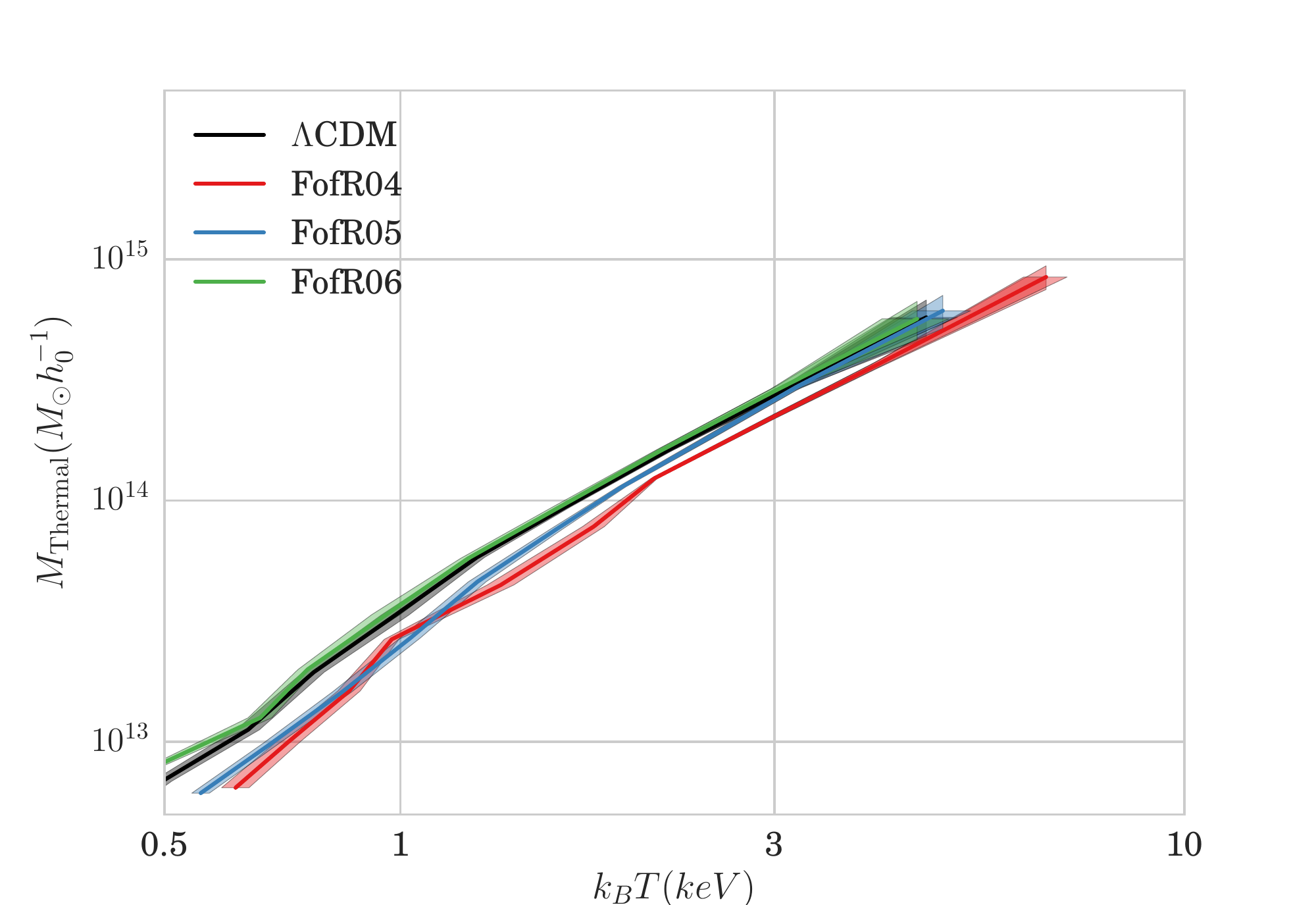}
	\vspace{-1 mm}
\caption{The mass-temperature relation for $\Lambda$CDM$_{\rm f(R)}$ and the $f(R)$-gravity models, for the stacked galaxy clusters.}
 \label{m-t-fofr}
 \end{figure}

In \fig{m-t-symmetron} and \fig{m-t-fofr} we present the mass-temperature relation for the symmetron- and $f(R)$-gravity models respectively. The broadening of the lines in the figures represents the standard deviation due to the stacking.

The massive, hot galaxy clusters, found at the top-right corner of the figures, show that all the models are indistinguishable from one another. However, the smaller, less hot clusters, show that the differences between the models get more pronounced the smaller and colder a cluster is. The differences between the models can clearly be distinguished when studying the smallest and coldest galaxy clusters.

A consequence of the minimum limit of dark matter particles described earlier is apparent when comparing the modified gravity models to the $Lambda$CDM models. The smallest clusters in the modified gravity models are noticeably larger and hotter than the smallest clusters in the $Lambda$CDM models. In the modified gravity models the temperature is larger than in $\Lambda$CDM \citep{hammami1, hammami2}, resulting in the thermal mass being noticeably larger in modified gravity models than in standard gravity. A comparison of how modified gravity theories affect the various masses (kinetic, lensing and thermal) can be found in \citet{max_hans_amir_david}.  


\fig{m-t-symmetron} show that the symmetron models deviate from $\Lambda$CDM in the order Sym D > Sym B > Sym C > Sym A, for the medium to low mass range. This demonstrate that the mass-temperature relation is more sensitive to the symmetry breaking criteria $a_{SSB}$ than the strength of the coupling $\beta$, similar to what was found in \citet{hammami1, hammami2}.

The $f(R)$ models show that the higher the coupling, the larger the deviations, with FofR04>FofR05>FofR06. However, unlike the symmetrons, we have a model that is permanently deviating from the $\Lambda$CDM value even for the largest clusters, namely that of FofR04. This is however not surprising, as FofR04 has long been ruled out as a viable candidate.

Due to the smaller box size employed in the $f(R)$ simulations the largest masses in this set of simulations are smaller than in the symmetron simulation set.

\begin{figure}
        \centering
        \includegraphics[width=0.45\textwidth]{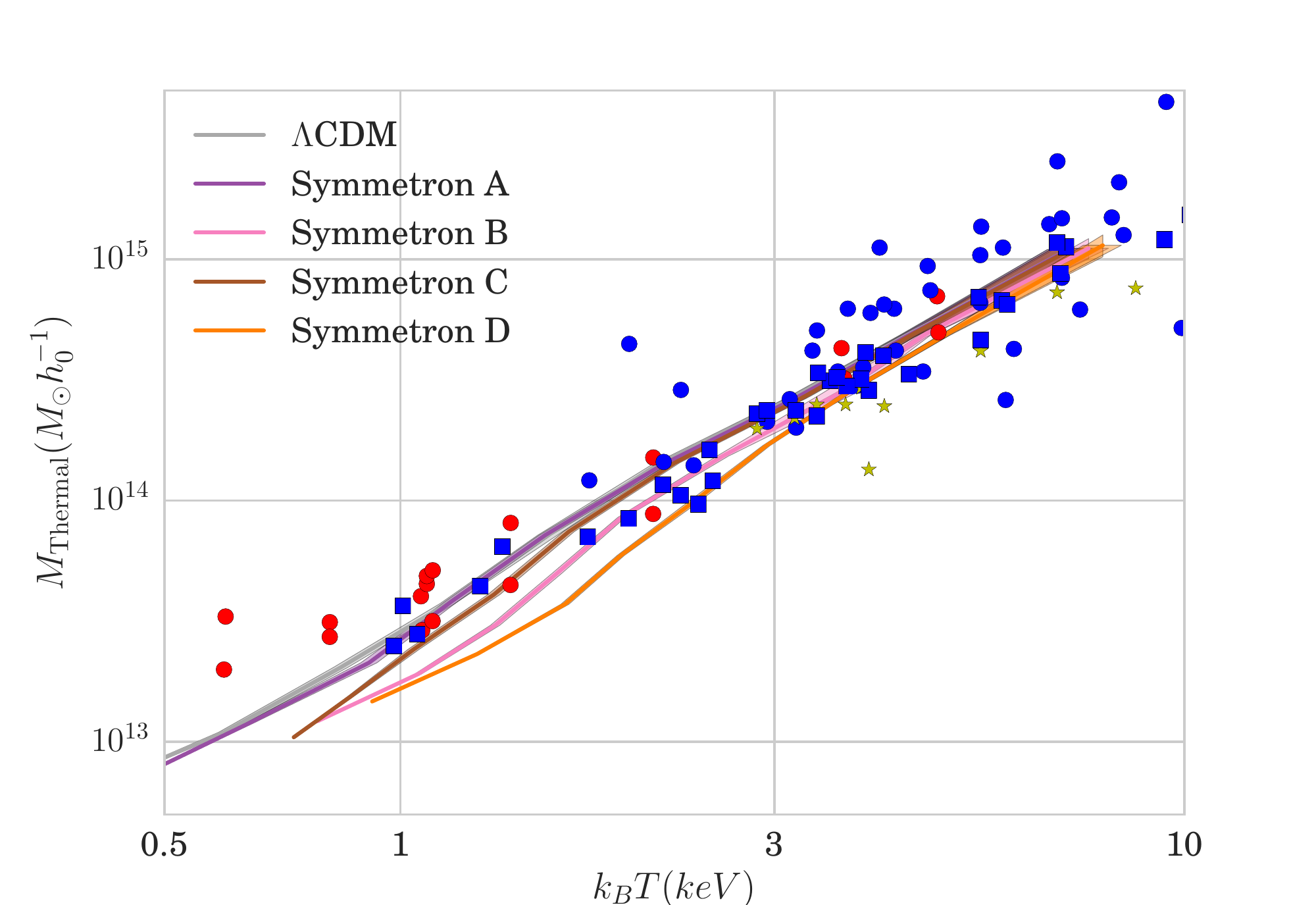}
	\vspace{-1 mm}
\caption{The mass-temperature data for $\Lambda$CDM$_{\rm S}$ and the symmetron models, for the stacked galaxy clusters.}
 \label{m-t-symmetron_obs}
 \end{figure}

\begin{figure}
        \centering
        \includegraphics[width=0.45\textwidth]{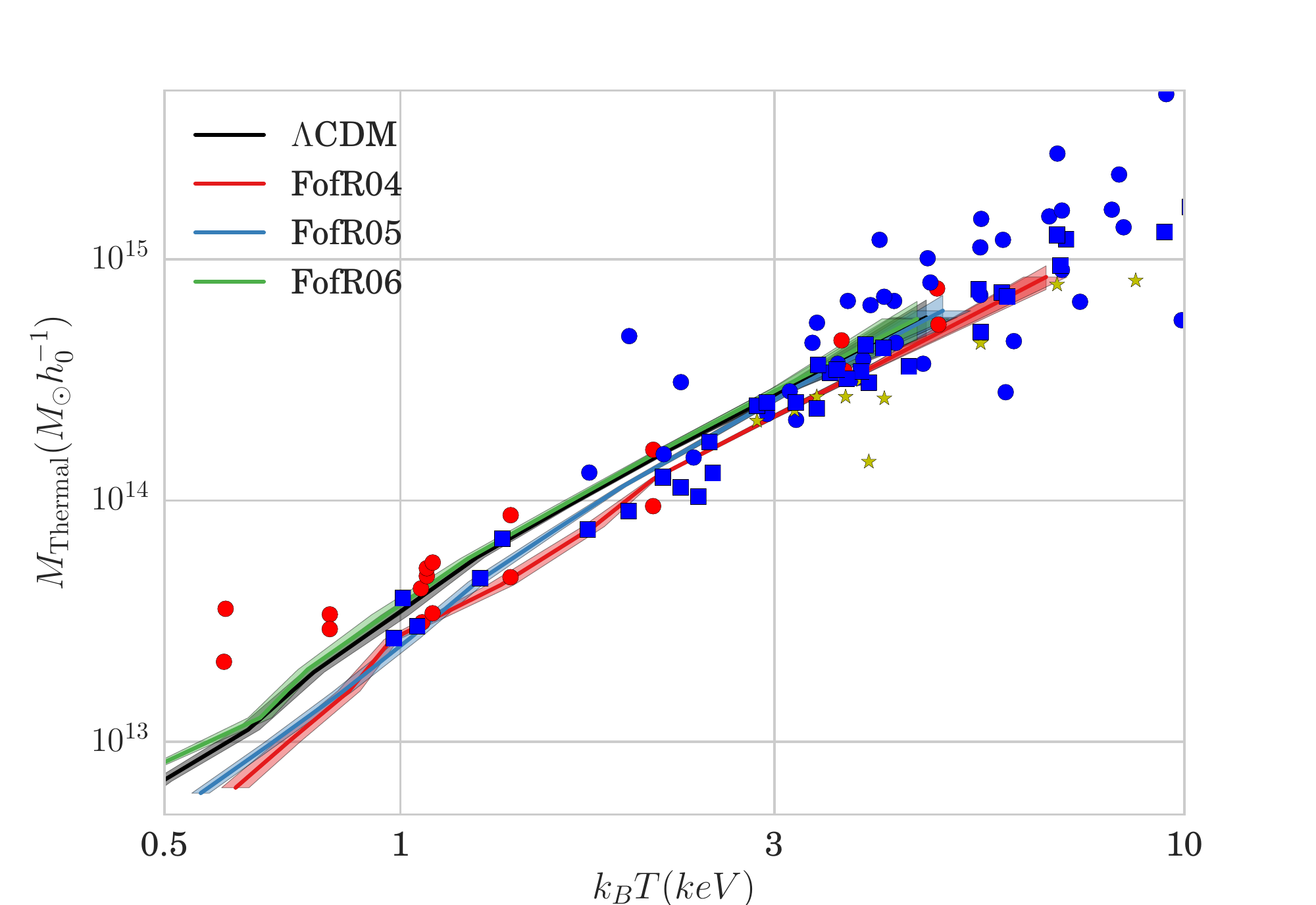}
	\vspace{-1 mm}
\caption{The mass-temperature data for $\Lambda$CDM$_{\rm f(R)}$ and the $f(R)$-gravity models, for the stacked galaxy clusters.}
 \label{m-t-fofr_obs}
 \end{figure}

In \fig{m-t-symmetron_obs} and \fig{m-t-fofr_obs} we present the observations from the literature as points overplotted on the previous two figures, for the symmetron- and $f(R)$-gravity models respectively. The red circles represent the data from \citet{X-ray_prop} and the blue points are data from \citet{PhD-thesis}. Additionally we plot the only spatially resolved data points that could be found fitting our criteria, curtsey of \citet{PhD-thesis}, as yellow stars. 

The observational data points clearly have a wide spread that encompasses all of our models at the largest masses. At the very lowest masses the observations have a higher value than any of our simulated models, however the number of data points in this region is also low.

The data using spatially resolved observations seem to have a lower mass-temperature value than all of our models, both standard and modified gravity. This has however been pointed out in the literature and our source of the data \citet{PhD-thesis}.

\fig{m-t-symmetron_obs} show that all models are consistent with observations at the large mass range, however at the medium-to-low end we note that the data is only consistent with standard gravity and Sym A. Sym B, Sym C and Sym D all fall substantially far below the observed data point, a clue that the mass-temperature relation could be a prime candidate for excluding modified gravity models. 

\fig{m-t-fofr_obs} show the same behaviour, but here all models remain consistent with observations for a wider mass-range than the symmetrons. First at temperatures below 1 keV can we point to FofR04 and FofR05 no longer being consistent with the observations.

\begin{figure}
        \centering
        \includegraphics[width=0.45\textwidth]{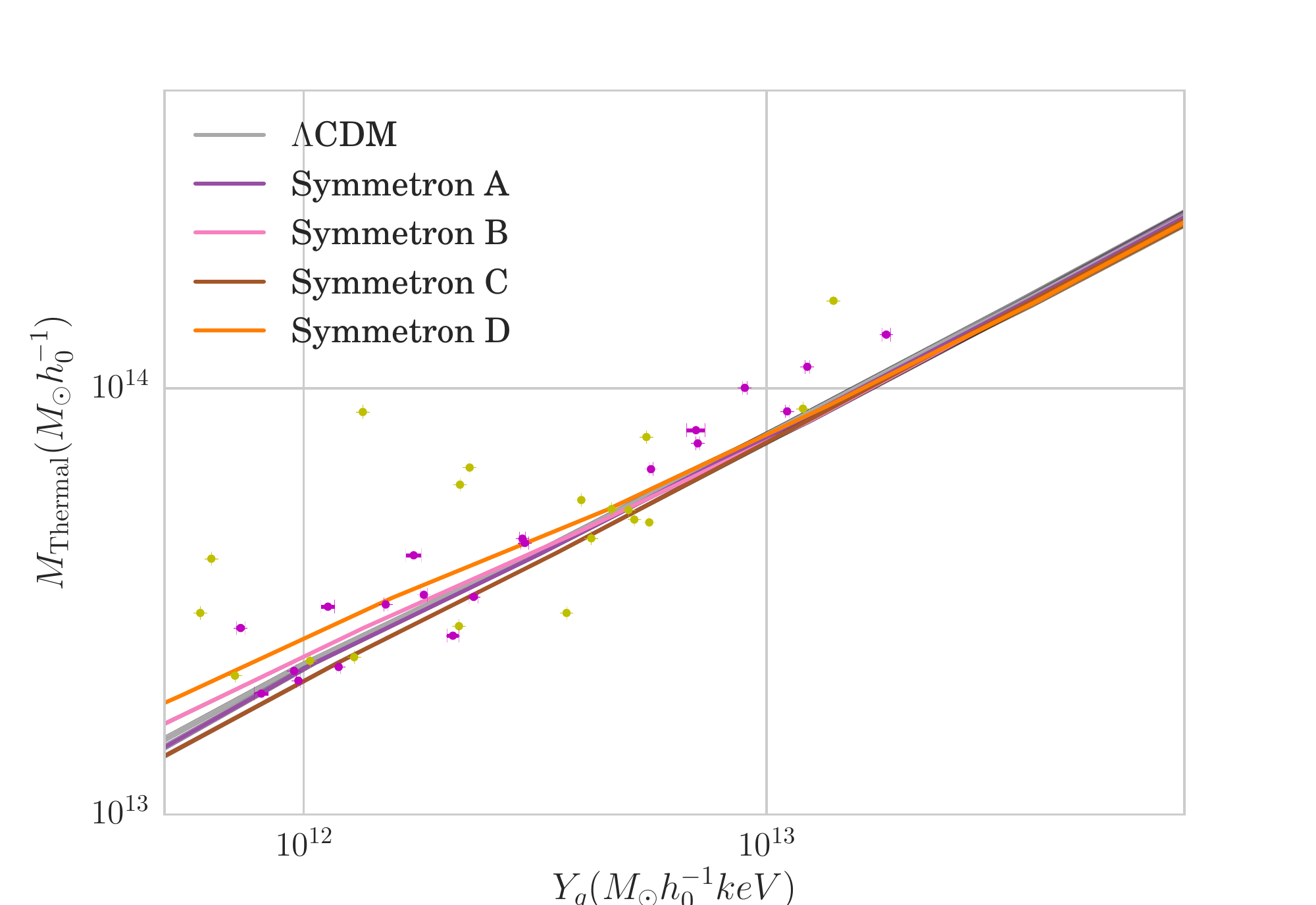}
	\vspace{-1 mm}
\caption{The mass-$Y_X$ data for $\Lambda$CDM$_{\rm S}$ and the symmetron models, for the stacked galaxy clusters.}
 \label{Y-t-symmetron_obs}
 \end{figure}

\begin{figure}
        \centering
        \includegraphics[width=0.45\textwidth]{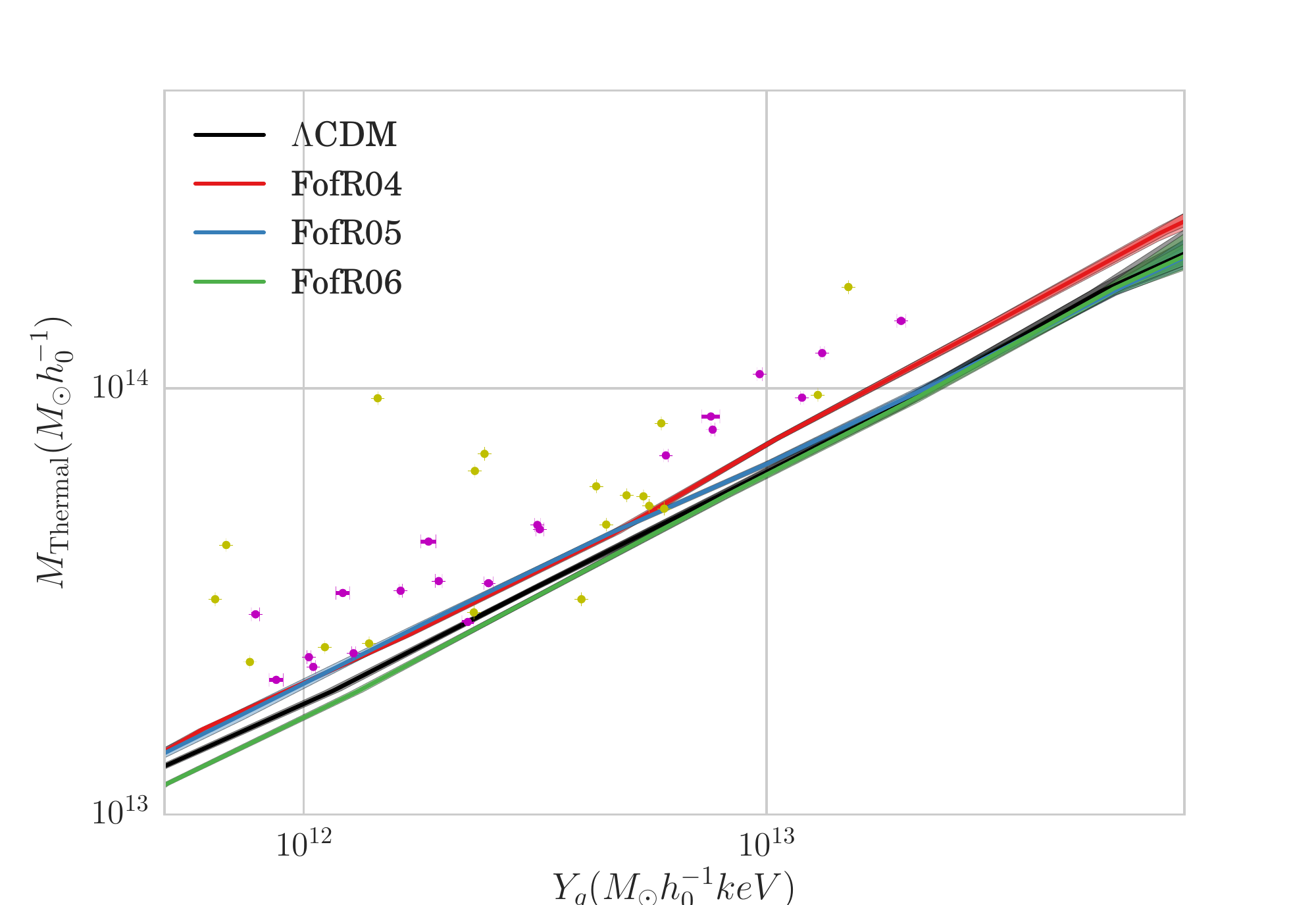}
	\vspace{-1 mm}
\caption{The M-$Y_X$ data for $\Lambda$CDM$_{\rm f(R)}$ and the $f(R)$-gravity models, for the stacked galaxy clusters.}
 \label{Y-t-fofr_obs}
 \end{figure}

In \fig{Y-t-symmetron_obs} and \fig{Y-t-fofr_obs} we present the proxy mass indicator relation M-$Y_X$ for the symmetron- and $f(R)$-gravity models respectively, while also including the observations from the literature as points. The yellow circles represent the data from \citet{Eckmiller11} and the magenta triangles are data from \citet{Lovisari15}.

It is clear that, although less sensitive to systematics, the mass-$Y_X$ relation is less suited for probing modified gravity theories than the mass-temperature relation. Scaling the temperature with the mass of the gas has the effect of diminishing the variations between the various models. This seems to tell us that the temperature and mass of the gas deviate in opposite ways, and when combined negate each other.

\begin{figure}
        \centering
        \includegraphics[width=0.45\textwidth]{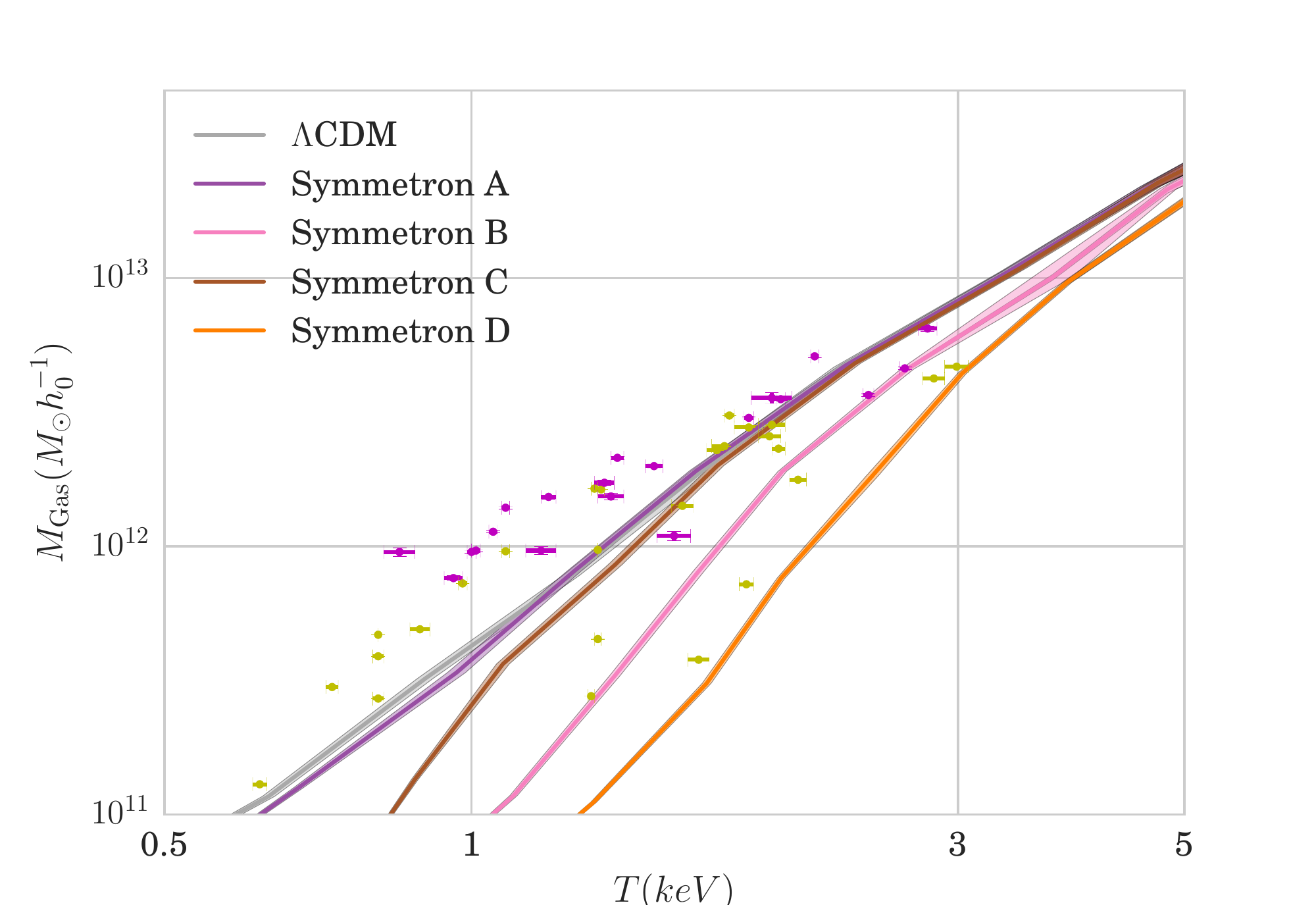}
	\vspace{-1 mm}
\caption{The gas mass-temperature data for $\Lambda$CDM$_{\rm S}$ and the symmetron models, for the stacked galaxy clusters.}
 \label{gas-symmetron_obs}
 \end{figure}

\begin{figure}
        \centering
        \includegraphics[width=0.45\textwidth]{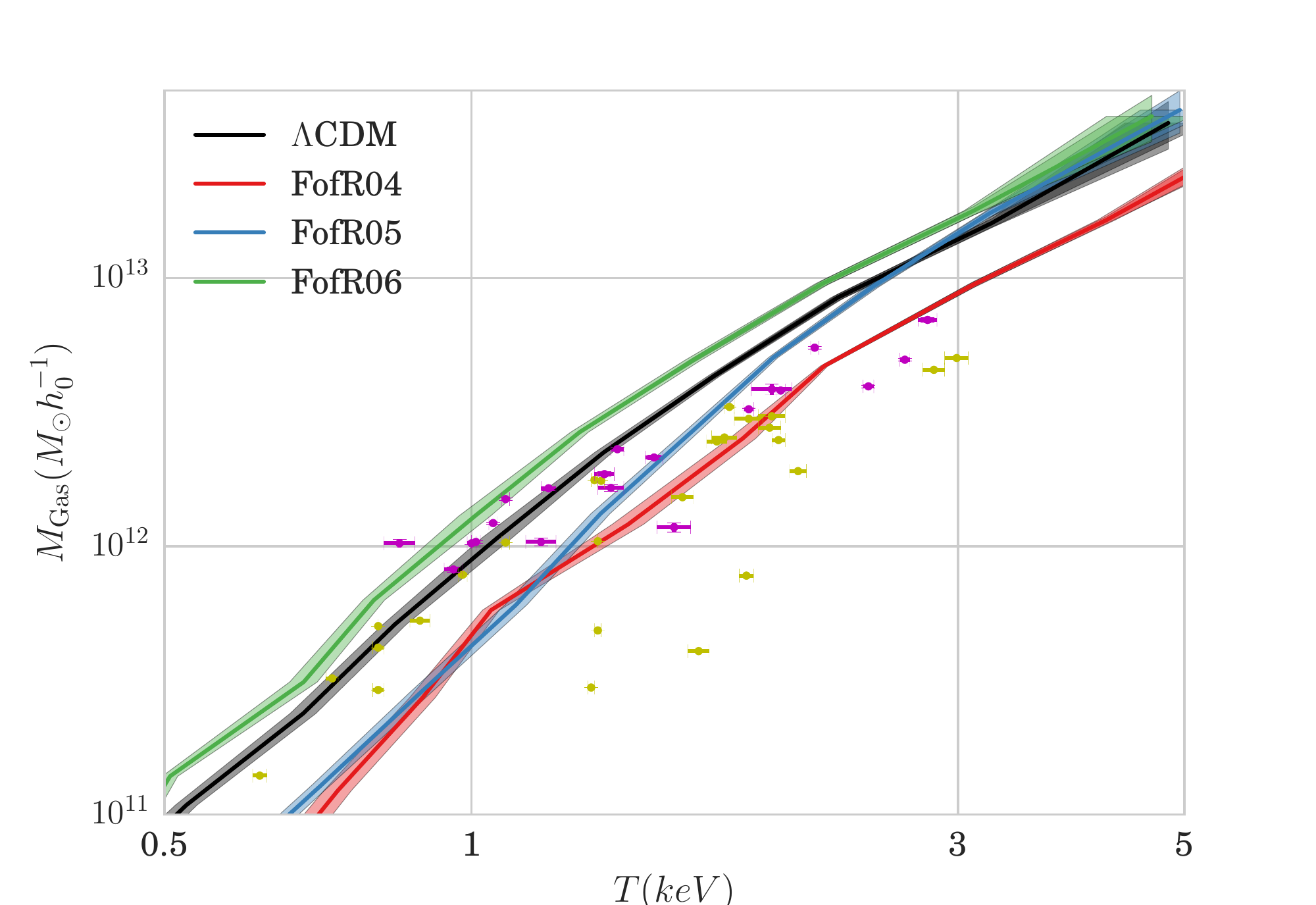}
	\vspace{-1 mm}
\caption{The gas mass-temperature data for $\Lambda$CDM$_{\rm f(R)}$ and the $f(R)$-gravity models, for the stacked galaxy clusters.}
 \label{gas-fofr_obs}
 \end{figure}

The compelling reason for using this mass proxy is that the gas of the mass is an easier observable than the thermal mass, which require several assumptions as detailed above. We therefore perform the analysis once more for the mass of the gas. In \fig{gas-symmetron_obs} and \fig{gas-fofr_obs} we present the mass-temperature relation for the gas mass instead of the thermal mass, with observations from the same sources as for the proxy.

The mass of the gas and the temperature does not follow a power law as clearly as the thermal mass and temperature is. However, the relation show much larger variations between the models than the two previous relations did. For the least massive clusters we can note as much as an order of magnitude difference between $\Lambda$CDM$_{\rm S}$ and Sym D. Other than the size of the deviations, the models deviate in the same order as before. 

\subsection{Analytical fit}
Theory predicts that the thermal mass and temperature should follow a power law function of the form of \refeq{theoretical_mt}, it is therefore of interest to see if we can fit a power law relation between the mass and temperature to our simulation results.

We perform a non-linear least squares fit, using the Levenberg-Marquardt algorithm \citep{Levenberg1944}, through the scipy Python package \citep{scipy} to fit the data with a power law function  
\begin{align}
M =  aT^b.
\end{align}

The fit analysis is run on the mass-temperature relation using the stacked galaxy clusters, to avoid skewing the results with the outliers. The results for $b$ and its corresponding standard deviations are presented in \reftable{fitt}.

\begin{table}
 \begin{center}
\caption{The result from fitting a power law to our simulated mass-temperature relation}
\label{params_tab}
  \begin{tabular}{lcc}\hline
  Models  & $a$ & $b$ \\ \hline
  $\Lambda$CDM$_{\rm S}$ & $0.360\pm 0.014$ & $1.742\pm 0.014$  \\ 
  $\Lambda$CDM$_{\rm f(R)}$ & $0.405\pm 0.025$ & $1.720\pm 0.025$  \\ 
  Sym A  & $0.347\pm 0.010$ & $1.760\pm 0.014$ \\ 
  Sym B  & $0.270\pm 0.021$ & $1.843\pm 0.038$ \\ 
  Sym C  & $0.323\pm 0.016$ & $1.794\pm 0.026$ \\ 
  Sym D  & $0.213\pm 0.015$ & $1.933\pm 0.035$ \\ 
  FofR04 & $0.325\pm 0.016$ & $1.718\pm 0.027$ \\ 
  FofR05 & $0.347\pm 0.019$ & $1.806\pm 0.036$ \\ 
  FofR06 & $0.421\pm 0.016$ & $1.717\pm 0.028$ \\ \hline\hline
  \end{tabular}
 \end{center}
\end{table}

The power law fit for  $\Lambda$CDM$_{\rm S}$and  $\Lambda$CDM$_{\rm f(R)}$ differ, however both are consistent with one another when considering the standard deviation. The amplitude is expected to differ due to the difference in box size of the simulation and general differences in masses and temperatures that follow.

The largest deviations from the  $\Lambda$CDM fits are found in the models Sym D and FofR05, while only FofR04 and FofR06 are consistent with $\Lambda$CDM$_{\rm f(R)}$.

The symmetron best fits follow a pattern of an increasing exponent and decreasing amplitude for a decreasing symmetry breaking criteria. With only one model with a coupling other than unity, we cannot, as of now, discern the effect the strength of the coupling has on the best fit. We can estimate a mass-temperature relation correlated with the symmetry breaking criteria $a_{SSB}$f as 
\begin{align}
M_{\rm Symmetron} = 0.600 a_{SSB}^{0.681}\times T^{1.600 a_{SSB}^{-0.134}}.
\end{align}
This relation assumes that $\beta=1$. The relation proves to have a 2.4\% accuracy in the amplitude and a 0.3\% accuracy in the exponent. Inserting $a_{SSB}=a_0 = 1$ into the relation above we find that the lower limit power law for the symmetron model is $T^{1.6}$, for models where $\lambda_0=1$ and $\beta=1$.

We are unable to find a direct correlation between the power and value of $f_{R0}$, however the amplitude is increasing with a decreasing $f_{R0}$. For both FofR04 and FofR06 the exponent is consistent with $\Lambda$CDM$_{\rm f(R)}$, with FofR05 having a larger power. By assuming that the power of FofR05 is a curiosity and that the power of the $f(R)$ mass-temperature relation is the same as in $\Lambda$CDM$_{\rm f(R)}$, we can construct a relation between $f_{R0}$ and the amplitude as
\begin{align}
M_{\rm F(R)} = 0.184f_{R0}^{-0.582}\times T^{\Lambda \rm CDM},
\end{align}
with a 5\% accuracy.

\subsection{Universal and non-universal coupling}
In our previous work \citep{hammami2} we studied the effect of having two different couplings to matter, one for baryons $\beta_{gas}$ and one for dark matter $\beta_{DM}$. We repeat part of the analysis above for the Sym B model, now with a wide range of various coupling combinations as shown in \reftable{tab:mgparam}
\begin{table}
 \begin{center}
\caption{Various couplings based on the Sym B model.\label{tab:mgparam}}
  \begin{tabular}{ccc}\hline
  Configuration & $\beta_{DM}$ & $\beta_{gas}$ \\ \hline
  \BothOne & 1.0 & 1.0\\
  \BothTen & 10 & 10\\
  \BothPointOne & 0.1 & 0.1\\
  \DMTen & 10 & 1.0\\
  \GasTen & 1.0 & 10\\
  \DMPointOne & 0.1 & 1.0\\
  \GasPointOne & 1.0 & 0.1\\
  \end{tabular}
 \end{center}
\end{table}

Our hope is to find some signature that can distinguish models with universal coupling from models with non-universal coupling. If one were to find traces of a non-universal coupling in observations this would essentially a breaking of the equivalence principle.

\begin{figure*}
\hspace{3cm}Universal \hspace{5cm}\;Non-universal\\
        \centering
        \includegraphics[width=0.4\textwidth]{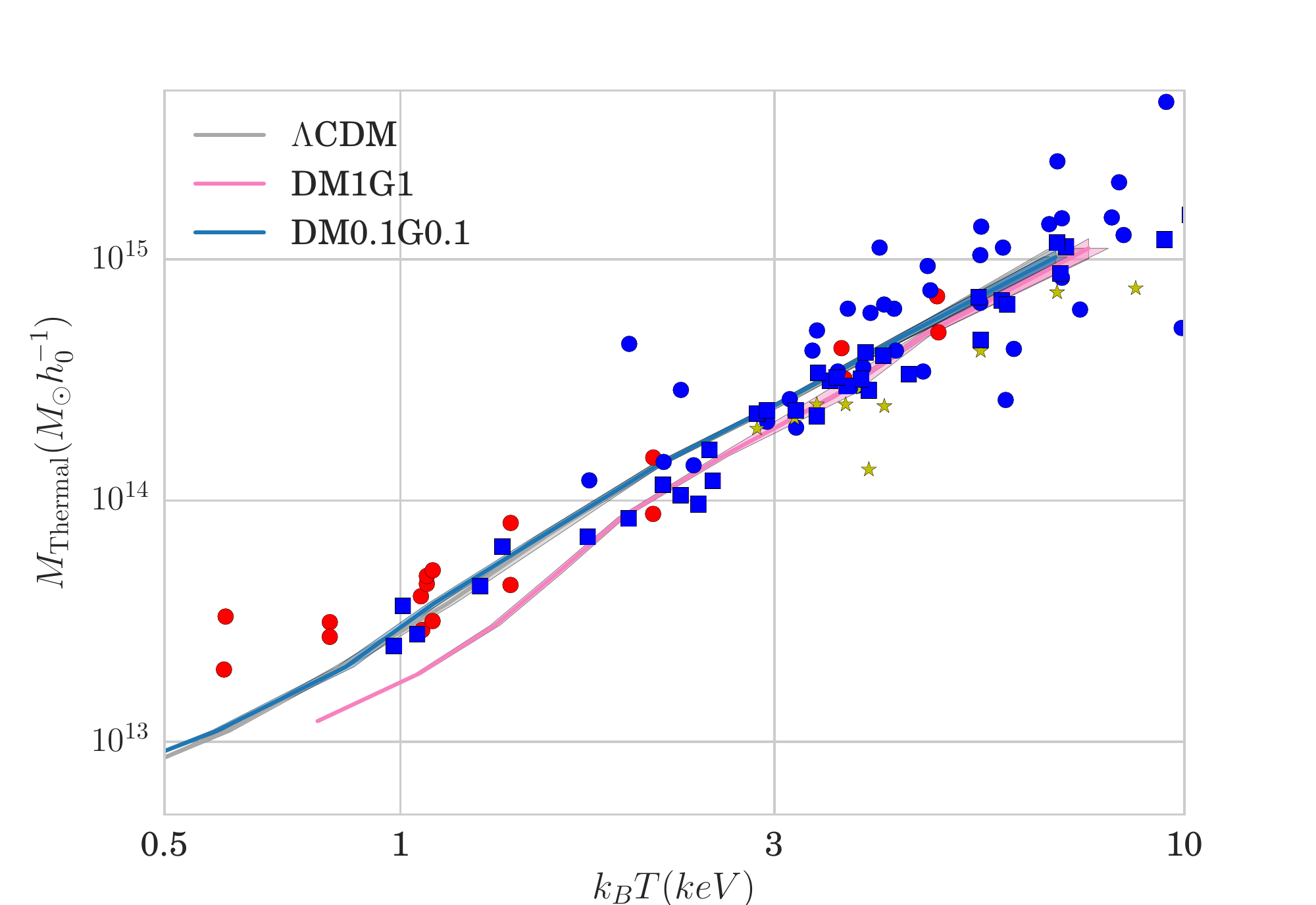}
	\hspace{-5 mm}
        \includegraphics[width=0.4\textwidth]{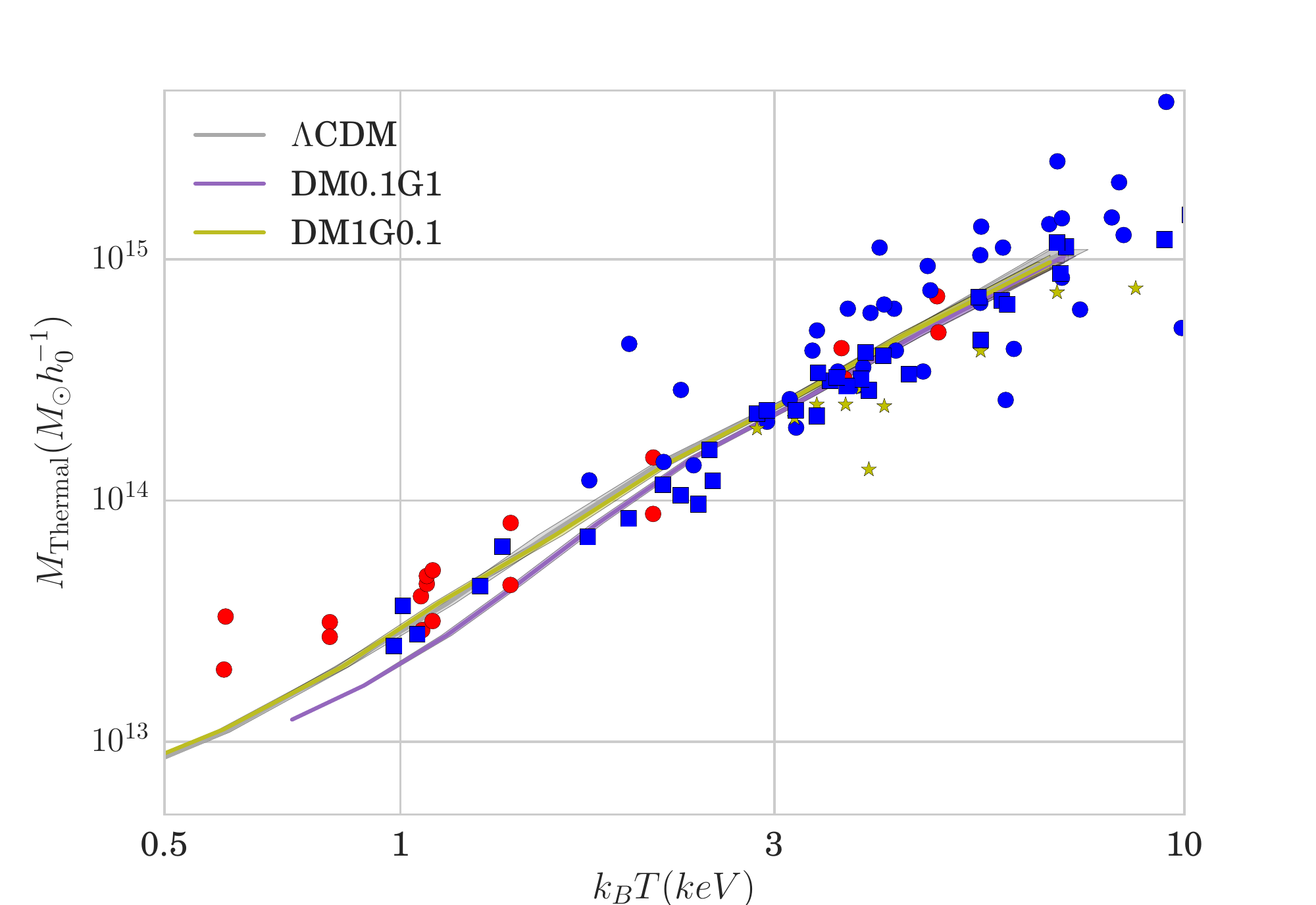}

        \includegraphics[width=0.4\textwidth]{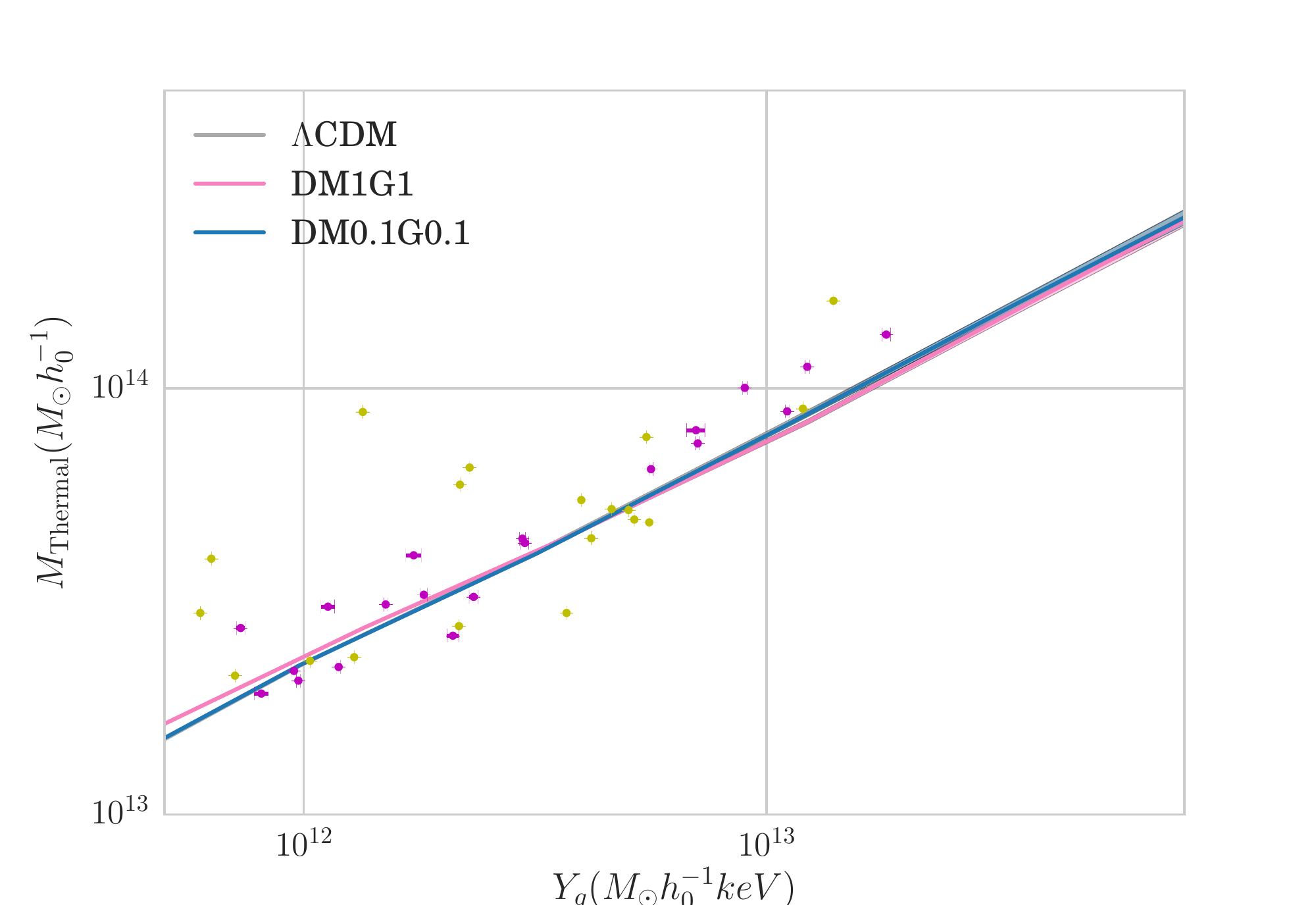}
	\hspace{-5 mm}
        \includegraphics[width=0.4\textwidth]{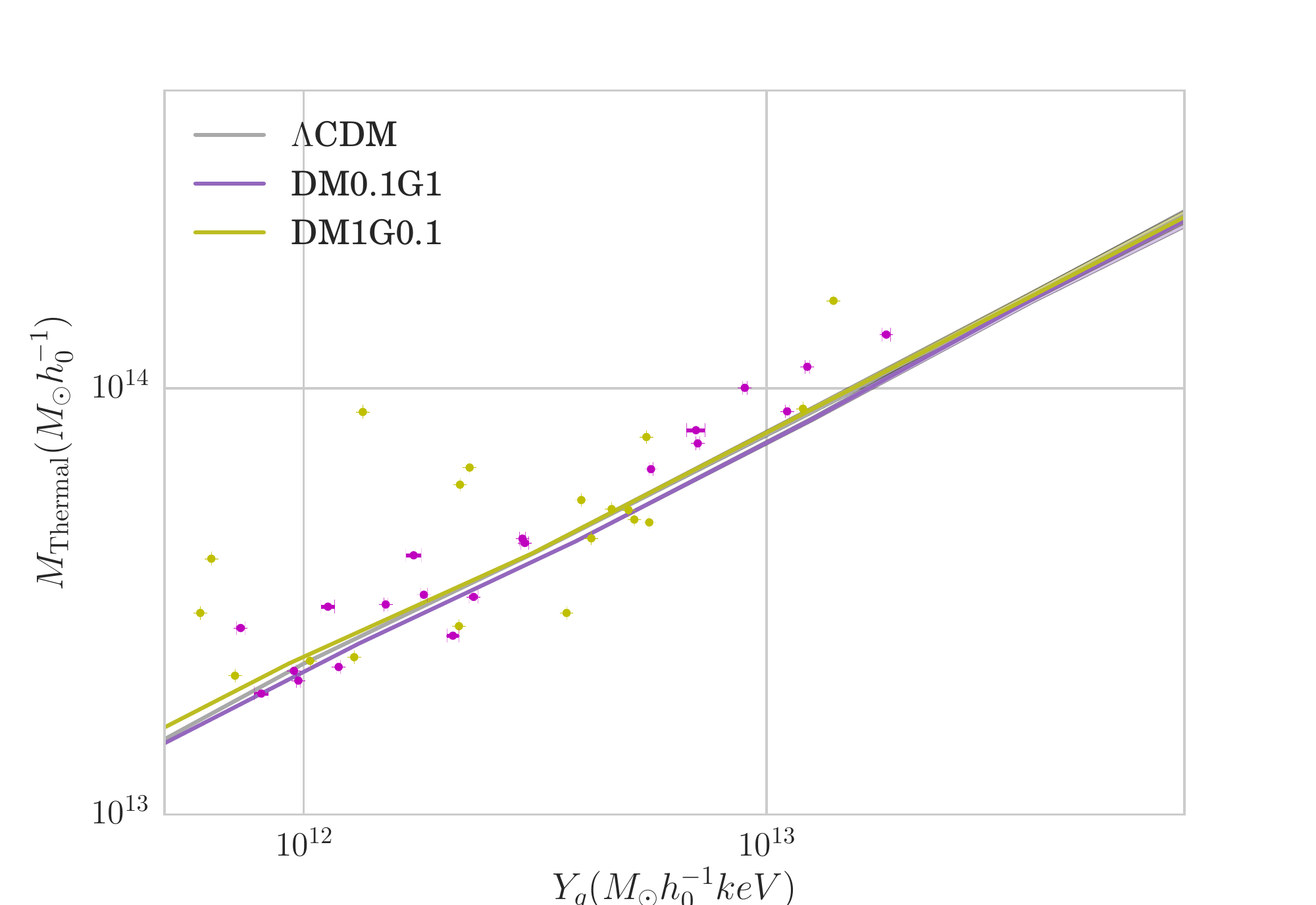}

        \includegraphics[width=0.4\textwidth]{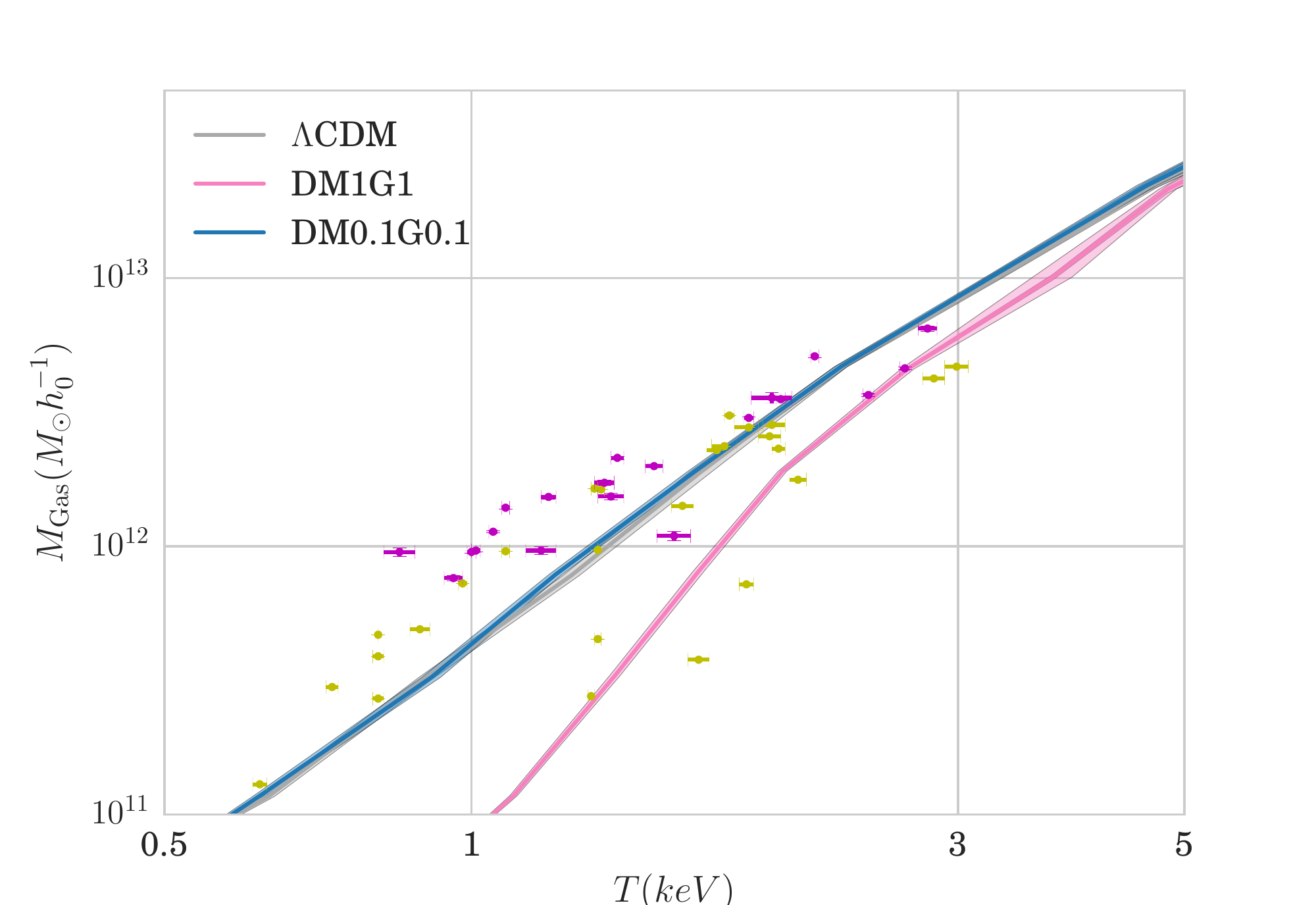}
	\hspace{-5 mm}
        \includegraphics[width=0.4\textwidth]{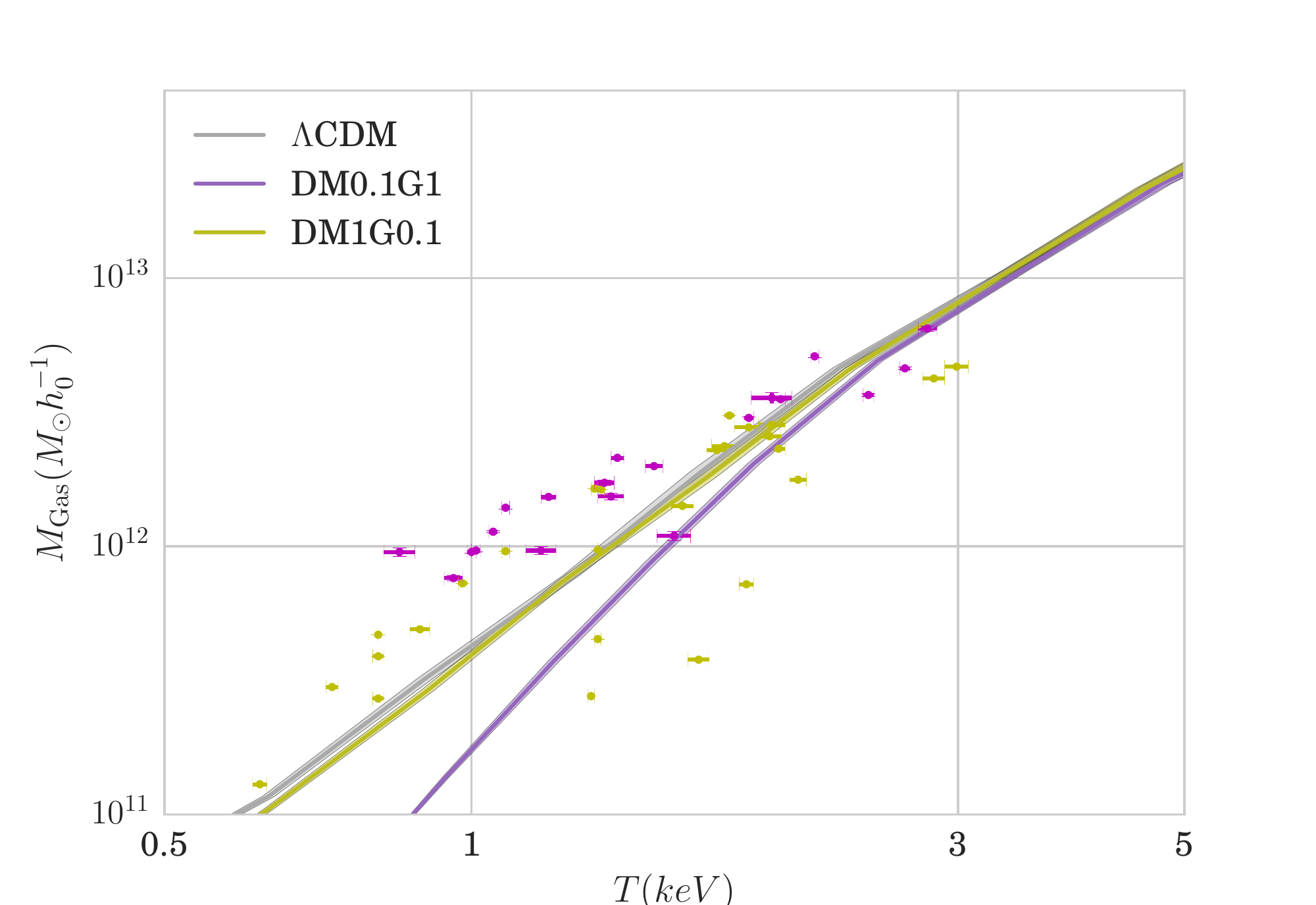}
	\vspace{-1 mm}
\caption{The top figures show the thermal mass-temperature ratio, the middle figures show the mass-$Y_X$ relation and the bottom figures show the gas mass-temperature relation, for universally and non-universally coupled models respectively.}\label{non-mass}
\end{figure*}

In \fig{non-mass} we present the thermal mass-temperature relation, the mass-$Y_X$ relation and gas mass-temperature relation for both the universal and non-universal models.  

Once again we see that we have a hard time distinguishing the models when studying the mass-$Y_X$ relation, a better time distinguishing models for the thermal mass-temperature relation and a very easy time distinguishing models in the gas mass-temperature relation.

Models where the gas is minimally coupled to the scalar field show very little deviations from the standard gravity mass-temperature relation. This effect is seen in all three types of mass-temperature relations. This finding is in agreement with results in \cite{hammami2}, where we noted that deviations in the temperature profiles were not noticeably different from standard gravity for a minimally coupled gas. As all the mass estimates above use either the gas or the temperature to estimate the mass, a minimally coupled gas will result in a mass that is also indistinguishable from the mass in standard gravity. 

No signatures that can identify universal coupling from non-universal coupling can be found.

\section{Conclusions}
\label{conclusions_sec}
In this paper we have shown that the mass-temperature relation can be a prime candidate for testing modified gravity theories against observations. 

The strength of studying the mass-temperature relation is that the modified gravity models can easily be distinguished from the $\Lambda$CDM reference values, masses below $M=5\times10^{14}$ M$_{\odot}h^{-1}$ and temperatures below $k_BT=1$ keV, the best-fit analysis returns a power-law very unlike the theoretical $M\propto T^{3/2}$, there already exist a framework for observing the mass-temperature relation and that there exist large quantities of data available.

Unfortunately the amount of observations available in this mass-temperature range is sparse, with the majority of the observations being for massive, hot galaxy clusters where all models are indistinguishable from one another. However with major surveys such as Chandra and XMM Newton the data is available, and only need to be assembled.

Alternatives to the standard thermal mass-temperature relation were explored, by using the mass proxy $Y_X$ as well as the mass of the gas directly. Both of these suffer less observational systematics than the thermal mass observations as the gas can be directly observed. The mass proxy unfortunately diminished the deviations that allowed us to easily distinguish the models. With the gas mass-temperature relation however the models could be distinguished to an even greater degree. We therefore propose that  the gas mass-temperature relation can be an even stronger candidate than the standard thermal mass-temperature relation.

It is imperative that the observers and the theoreticians running simulations communicate, at the moment the observables are presented in a wide range of variable ways. A consensus for what radius $r_{\Delta c}$ to perform the measurements at and whether to use a fitting model for the cluster profiles or actual spatially resolved profiles need to be reached.

We showed that, for the symmetron models, the thermal mass-temperature relation is strongly sensitive to the symmetry breaking criteria $a_{SSB}$ with a smaller dependency on the strength of the coupling to the scalar field. The analytical fit for the symmetron models also showed a high sensitivity to the symmetry breaking criteria, and we constructed a mass-temperature relation with $a_{SSB}$ as an input parameter, for all symmetron models with a coupling factor $\beta=1$.

While the power of the $f(R)$ mass-temperature relations did not appear to follow any relation to the choice of $f_{R0}$, we are able to present a relation for the amplitude of the mass-temperature relation and the input parameter $f_{R0}$. 

Numerous pitfalls still exist for using the mass-temperature relation as a primer on modified gravity theories, such as the assumption of galaxy clusters being in hydrostatic equilibrium, the vast number of uncertainties related to observational astronomy and numerical uncertainties. However, armed with the information presented in this paper a renewed focus on studying and understanding the mass-temperature relation in future studies should be warranted. 

Furthermore, if the gas is minimally coupled to the scalar field then the temperature of the gas is unchanged, and therefore the thermal mass shows no deviations from general relativity. This means that even if the dark matter is coupled to the scalar field, the mass-temperature relation may still be indistuingshable from standard gravity. The mass-temperature relation is thus a poor probe of gravity for models where the baryons are minimally coupled to the scalar field.
 
\section{Acknowledgements}
The authors thank The Research Council of Norway for funding and the NOTUR facilities for the Computational resources. 

\bibliography{mass_temp}

\end{document}